\documentclass[systems,article,submit,pdftex,moreauthors]{Definitions/mdpi}
\usepackage{comment}

\firstpage{1} 
\pubvolume{0}
\issuenum{0}
\articlenumber{0}
\pubyear{2025}
\copyrightyear{2025}
\datereceived{ } 
\daterevised{ } % Comment out if no revised date
\dateaccepted{ } 
\datepublished{ } 
\hreflink{https://doi.org/} % If needed use \linebreak
\Title{OSCILLATOR CHAIN MODEL FOR MULTI-CONTOUR SYSTEMS WITH PRIORITY IN CONFLICT RESOLUTION}
\Author{Ihor Lubashevsky  $^{1,2}$, Marina Yashina $^{2}$, and Vasily Lubashevskiy $^{3,*}$}
\AuthorNames{Ihor Lubashevsky, Marina Yashina, and Vasily Lubashevskiy}

\AuthorCitation{Lubashevsky, I.; Yashina M. ; Lubashevskiy V.}
	
\address{%
	$^{1}$ \quad HSE University, Tikhonov Moscow Institute of Electronics and Mathematics, 34 Tallinskaya Ulitsa, Moscow 123458,  Russia; ilubashevskii@hse.ru\\
	$^{2}$ \quad Moscow Automobile and Road Construction Technical University, Department of Mathematics, Leningradsky~Ave. 64, Moscow 125319, Russia; mv.yashina@madi.ru\\
	$^{3}$ \quad Tokyo International University, Institute for International Strategy, 4 Chome-42-31 Higashiikebukuro, Toshima, Tokyo 170-0013, Japan; vlubashe@tiu.ac.jp}
	
\corres{Corresponding author}
	
\nolinenumbers
\abstract{
We propose a novel model of oscillatory chains that generalizes the contour discrete model of Buslaev nets. The model offers a continuous description of conflicts in system dynamics, interpreted as interactions between neighboring oscillators when their phases lie within defined interaction sectors. The size of the interaction sector can be seen as a measure of vehicle density within clusters moving along contours. The model assumes that oscillators can synchronize their dynamics, using concepts inherited from the Kuramoto model, which effectively accounts for the discrete state effects observed in Buslaev nets. The governing equation for oscillator dynamics incorporates four key factors: deceleration caused by conflicts with neighboring oscillators and the synchronization process, which induces additional acceleration or deceleration. Numerical analysis shows that the system exhibits both familiar properties from classic Buslaev nets, such as metastable synchronization, and novel behaviors, including phase transitions as the interaction sector size changes. 
}

	
\begin{document}

\section{Introduction}

A variety of self-organization phenomena in traffic flow on complex networks can be described using the contour network approach, first proposed in \cite{kozlov2013synergy} and commonly referred to as Buslaev nets. Such a net consists of a set of contours that represent the cyclic motion of individual dynamical systems through a sequence of states. These systems interact at shared points of adjacent contours, which serve as the nodes of the Buslaev nets. This interaction locally decelerates system dynamics and can ultimately lead to a complete cessation of motion. Asymmetry in motion deceleration is analyzed in terms of priority-based conflict resolution.

When Buslaev nets are used to structurally represent complex urban street networks, the dynamical system associated with each contour is interpreted as a cluster of vehicles moving along it. Vehicles traveling on different contours cannot pass through the nodes simultaneously, which forms the basis of conflict dynamics and contributes to the slowing of system dynamics. Differences in street hierarchy are incorporated through priority-based conflict resolution.

In \cite{kozlov2013synergy,Bugaev2021}, a one-dimensional symmetric network, referred to as a closed chain, is studied. To analyze the qualitative behavior of dynamical systems of this type, the authors introduced the concept of a velocity spectrum and defined a limit cycle as a submanifold of the state space where a fixed-speed regime is realized. In \cite{kozlov2013synergy,Buslaev2013}, a two-dimensional network called a "chainmail" was considered. Each circuit shares a common point---a node connecting two adjacent contours. A moving segment cluster propagates through each circuit at a constant velocity in the absence of delays. Delays in cluster movement arise due to interactions at the common node. The spectrum of average velocities and the conditions for self-organization were derived.
In \cite{Buslaev2018,Yashina2019,Yashina2021,Bugaev2023}, an open chain was studied, in which the leftmost and rightmost elements of the network each have only one adjacent circuit. In \cite{Myshkis2020,Yashina2020}, a two-circuit system with one or two nodes, respectively, was analyzed.
	
In the present paper, we propose a novel approach to describing cooperative phenomena in Buslaev nets, which can be regarded as analogous to the Kuramoto model for synchronization phenomena (see \cite{Acebron2005} for a review). To illustrate the core idea of the proposed approach, we restrict our analysis to a closed chain of contours.

\section{Model}

\subsection{Conflict description}

\begin{figure}
	\begin{center}
		\includegraphics[width=1\textwidth]{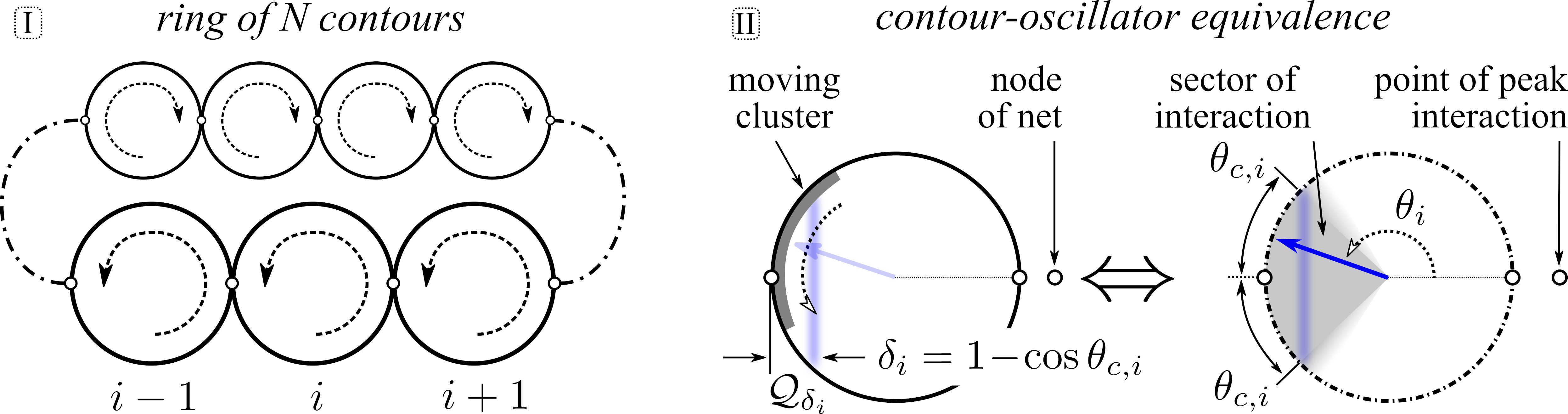}
	\end{center}
	\caption{The analyzed closed chain of unit-radius contours (I) and the transformation of contours into oscillators (II). The left part of (II) illustrates a contour $i$ with a vehicle cluster moving along it. The cluster's size defines a neighborhood $\mathcal{Q}_{\delta_i}$ around the corresponding node, where the cluster presence may cause a conflict with the motion of a vehicle cluster in the neighboring contour $i-1$. The blurred blue line represents the boundary of the fuzzy neighborhood $\mathcal{Q}_{\delta_i}$, characterized by width $\delta_i$. The right part of (II) shows the equivalent representation of this cluster-contour arrangement in terms of oscillator $i$, depicted as the circle $e^{i\theta_i}$ in the complex plane. The phase $\theta_i$, indicated by the blue arrows, corresponds to the center of the moving cluster. The potential interaction sector with the neighboring oscillator, shown as a darkened region, is determined by the neighborhood $\mathcal{Q}_{\delta_i}$. As observed, the interaction sector size $\theta_{c,i}$ and the width $\delta_i$ of $\mathcal{Q}_{\delta_i}$ are related as $\delta_i = 1 -\cos\theta_{c,i}$, with $\theta_{c,i}$ serving as a measure of the vehicle cluster size.}
	\label{F:1}
\end{figure}

The analyzed system consists of a closed chain of $N$ unit-radius contours (Fig.~\ref{F:1}.\,I). Figure ~\ref{F:1}.\,II illustrates the proposed interpretation of a contour $i$ with a vehicle cluster moving along it as an oscillator $e^{i\theta_i}$, where the phase $\theta_i$ corresponds to the center of the vehicle cluster. The proposed model represents this Buslaev net as a corresponding network of such oscillators and, in the standard manner, treats $\cos\theta_i$ as the special position of oscillator $i$. 
The localization of the vehicle cluster near one of the contour nodes, which may lead to a conflict with a vehicle cluster moving along the neighboring contour, is described by introducing a sector of possible interaction between the given oscillator and its neighbor. The size of this sector---represented by the angle $\theta_c$---can, on the one hand, be interpreted as the size of the vehicle cluster. On the other hand, it defines the neighborhood of two critical points $\theta=0$ and $\theta=\pi$, corresponding to the network nodes where the interaction between neighboring oscillators reaches its maximum. The width of this neighborhood is estimated as $\delta_i = 1- \cos \theta_{c,i}$. We assume that oscillators may be characterized by individual values of $\delta_i$ and $\theta_{c,i}$ interrelated via the latter expression.

\begin{figure}
	\begin{center}
		\includegraphics[width=0.90\textwidth]{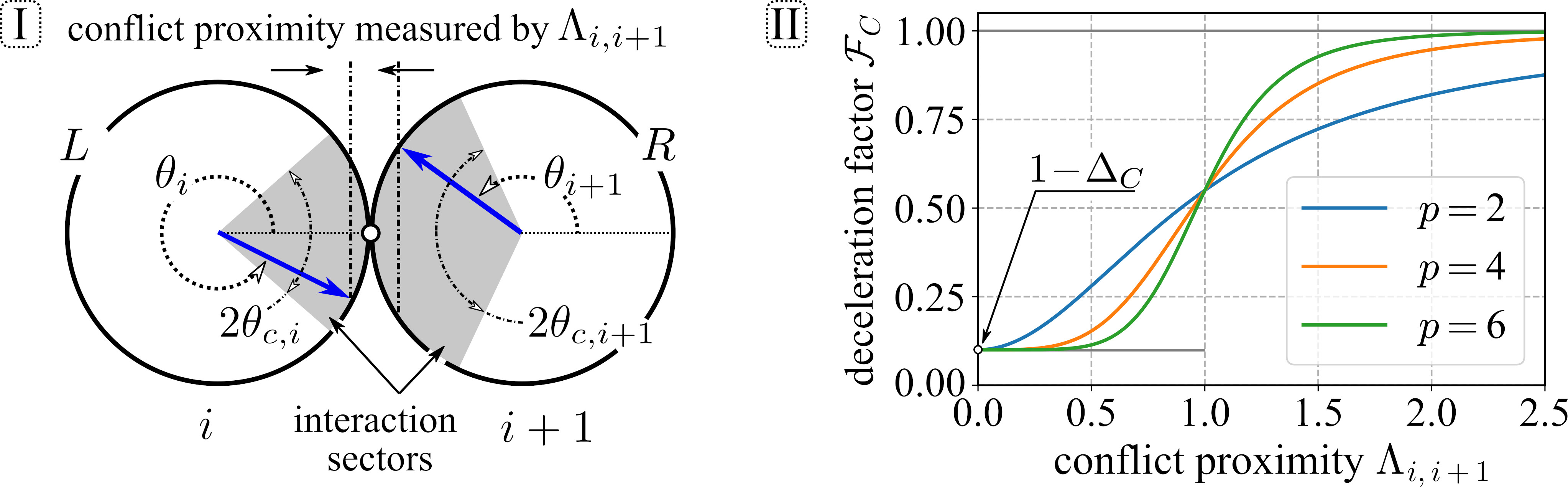}
	\end{center}
	\caption{The measure $\Lambda_{i,i+1}$ quantifying the proximity of an oscillator pair $\{i,i+1\}$ to the ultimate conflict configuration (I) and the form of the deceleration factor $\mathcal{F}_{L,R}(\theta_i,\theta_{i+1})$ as a function of $\Lambda_{i,i+1}$ for several values of its parameter $p$ (Eq.~\ref{eq:2}) (II). Blue arrows represent oscillators $i$ and $i+1$ as circles $e^{i\theta_i}$ and $e^{i\theta_{i+1}}$ on the complex plane, where $\theta_i$ and $\theta_{i+1}$ denote their phases. The indices $C=L$ and $C=R$ indicate the oscillator (left or right) whose motion deceleration is quantified by the factor $\mathcal{F}_{C}$, respectively. }
	\label{F:2}
\end{figure}

In the proposed model, the proximity of an oscillator pair $\{i,i+1\}$ to the ultimate conflict configuration, defined by $\theta_i \equiv 0 \pmod{2\pi}$ and $\theta_{i+1} \equiv \pi \pmod{2\pi}$,  is  quantified by a measure $\Lambda_{i,i+1}$ (Fig.~\ref{F:2}\,I) given by the expression
\begin{subequations}\label{eq:123}
\begin{equation}\label{eq:1a} 
	\Lambda_{i,i+1} = \bigg[\frac12\bigg(\frac{1 - \cos\theta_{i\phantom{,c}}}{1-\cos\theta_{c,i}}\bigg)^q + 
	\frac12\bigg(\frac{1 +\cos\theta_{i+1\phantom{,c}}}{1-\cos\theta_{c,i+1}}\bigg)^q\bigg]^{1/q}, 
\end{equation}
where the exponent $q\geq 1$. The coefficient $1/2$ in Eq.~\eqref{eq:1a} ensures that the measure $\Lambda_{i,i+1}$ takes the value 1 when both oscillators are at the boundary of their interaction sectors. In particular, for $q\to\infty$, we obtain 
\begin{equation}\label{eq:1b} 
	\Lambda_{i,i+1} = \max\bigg[\bigg(\frac{1 - \cos\theta_{i\phantom{,c}}}{1-\cos\theta_{c,i}}\bigg), 
	\bigg(\frac{1 +\cos\theta_{i+1\phantom{,c}}}{1-\cos\theta_{c,i+1}}\bigg)\bigg]. 
\end{equation}
In the present analysis, we restrict ourselves to the case $q=1$, where 
\begin{equation}\label{eq:1c} 
	\Lambda_{i,i+1} = \frac12\bigg(\frac{1 - \cos\theta_{i\phantom{,c}}}{1-\cos\theta_{c,i}}+ 
	\frac{1 +\cos\theta_{i+1\phantom{,c}}}{1-\cos\theta_{c,i+1}}\bigg) 
\end{equation}
and the mutual contribution of the oscillator pair $\{i,i+1\}$ to $\Lambda_{i,i+1}$  is most pronounced. 
\end{subequations}

The factors  $0\leq\mathcal{F}_{C}(\theta_i,\theta_{i+1})\leq 1$ for $C=L$ and $C=R$, which account for the slowing down of oscillator dynamics due to this conflict, are specified by the ansatz (Fig.~\ref{F:2}\,II): 
\begin{equation}\label{eq:2} 
	\mathcal{F}_{C}(\theta_i,\theta_{i+1}) = 1 -  \frac{\Delta_{C}}{1+\Lambda^p_{i,i+1}},
\end{equation}
Here the exponent $p>0$ is treated as a fixed parameter and the parameters $0\leq \Delta_{C}\leq 1$ determine the maximum degree of the deceleration effect, while the indices $C=L$ and $C=R$ specify which oscillator in the pair $\{i,i+1\}$ is being described. In other words, the factors $\mathcal{F}_{L}(\theta_i,\theta_{i+1})$ and $\mathcal{F}_{R}(\theta_i,\theta_{i+1})$ quantify the deceleration effect in the dynamics of oscillators $i$ and $i+1$, respectively.
The difference in the parameters $\Delta_{L}$ and $\Delta_{R}$ provides a mathematical description of the priority in resolving conflicts between vehicle clusters as they move through the nodes of Buslaev nets. 
In principle, the values of $\Delta_{C}$ may be specific to each oscillator.

\subsection{Oscillator synchronization}

As seen below, conflict resolution alone cannot lead to the emergence of ordered patterns in the oscillator arrangement. In discrete versions of the Buslaev nets, pattern formation arises due to the discreteness of vehicle cluster states. In the developed model, a dedicated synchronization mechanism is required. To address this, we turn to the well-known Kuramoto model, which describes synchronization phenomena (see, e.g., \cite{Acebron2005} for a review).

We assume that vehicle drivers attempt to avoid reaching a contour node when a neighboring vehicle cluster is passing through it. In terms of oscillator dynamics, this translates into a tendency for oscillator $i$ to synchronize its phase $\theta_i$ with the phase $\theta_{i+1}$ or $\theta_{i-1}$ of the neighboring oscillator. The physical implementation of this synchronization manifests as an acceleration or deceleration of the corresponding cluster motion, constrained by the flexibility of its current state. This effect is described by introducing a factor $\mathcal{S}_{C}(\theta_i,\theta_{i+1})$,  which, for the oscillator pair $\{i,i+1\}$, is given as follows:
\begin{equation}\label{eq:sync1} 
	\mathcal{S}_{C}(\theta_i,\theta_{i+1}) = \Big[1 + K_C \sin\big(\theta^*_{C}-\theta_{C}\big)\big],\quad\text{here}\quad \theta^*_{C}-\theta_{C} =
	\begin{cases}
		          \theta_{i+1} - \theta_{i}, & C=L\,,\\
		          \theta_{i} - \theta_{i+1}, & C=R \,.
	\end{cases}	                    
\end{equation}
The constants $K_C\geq 0$ ($C=L,R$) should satisfy the condition $K_C \leq 1$, since, in the contour model, vehicle clusters do not change their direction of motion; at worst, their motion may be blocked. It is natural to assume that a low priority in conflict resolution, i.e., high values of $\Delta_C$ indicating strong dynamic suppression, should encourage vehicle drivers to synchronize their states with the motion states of the neighboring vehicle cluster. 
In terms of oscillator dynamics, the stronger the rotational suppression of oscillator $i$ due to conflict with its neighboring oscillator $i+1$ (or $i-1$), the higher the constant $K_C$ should be. For example, the following relations:
\begin{equation}\label{eq:sync2}
	K_L = \kappa \Delta_L\,,\qquad K_R = \kappa \Delta_R\,,
\end{equation}	 
where $\kappa\leq 1$ is a constant, account for this aspect.

\subsection{Governing equation}

The resulting dynamics of these oscillators $\{e^{i\theta_i(t)}\}$  is assumed to be governed by the following equations:
\begin{equation}\label{eq:3} 
	\frac{d\theta_i}{dt} = \omega_i \mathcal{F}_R(\theta_{i-1},\theta_{i}) \mathcal{S}_{R}(\theta_{i-1},\theta_{i})  \mathcal{F}_L(\theta_{i},\theta_{i+1}) \mathcal{S}_{L}(\theta_i,\theta_{i+1})\, , 
\end{equation}
where $\omega_i$ is the individual rotation frequency of oscillator $i$ in the absence of conflicts. The product of the factors $\mathcal{F}_R(\theta_{i-1}, \theta_{i})$ and $\mathcal{F}_L(\theta_{i}, \theta_{i+1})$ accounts for the cumulative effect of interactions between oscillator $i$ and its neighboring oscillators $i-1$ and $i+1$. In other words, we assume that the motion of the vehicle cluster along contour $i$ can slow down due to conflicts with neighboring vehicle clusters as they simultaneously move through the nodes of contour $i$, provided that the size of the vehicle cluster is sufficiently large. The product of the factors $\mathcal{S}_R(\theta_{i-1}, \theta_{i})$ and $\mathcal{S}_L(\theta_{i}, \theta_{i+1})$ describes the synchronization of oscillator $i$ with its neighbors, with different priorities depending on the conflict strength. 

\section{Results}

In this paper, we present a preliminary investigation of the properties exhibited by the developed model. A ring of 501 oscillators (an odd number) was studied numerically using the ``odeint'' and ``solve\_ivp'' libraries from SciPy~1.15 for integrating ordinary differential equations, with the RK54, DOP853, and LSODA methods. It was found that ``odeint'' with RK54 and an integration step of 0.01 provides reliable results with minimal computation time. The total integration time $T$ was chosen in the range 
$T= 1000$ to 2000.

All oscillators were assumed to be identical in terms of the size of their interaction sectors, i.e., $\theta_{c,i} = \theta_c$. Consequently, the measure $\Lambda_{i,i+1}$ of conflict proximity can be calculated as 
\begin{equation}\label{eq:G}
	\Lambda_{i,i+1} = (2 - \cos\theta_i + \cos\theta_{i+1}) / G\,,
\end{equation}	
where $G = 2 (1-\cos\theta_c)$. In particular, $G=2$ corresponds to the case where the left and right interaction sectors of each oscillator begin to overlap. By varying 
$G$, we can effectively study the impact of vehicle density.

\begin{figure}[t]
	\begin{center}
		\includegraphics[width=\textwidth]{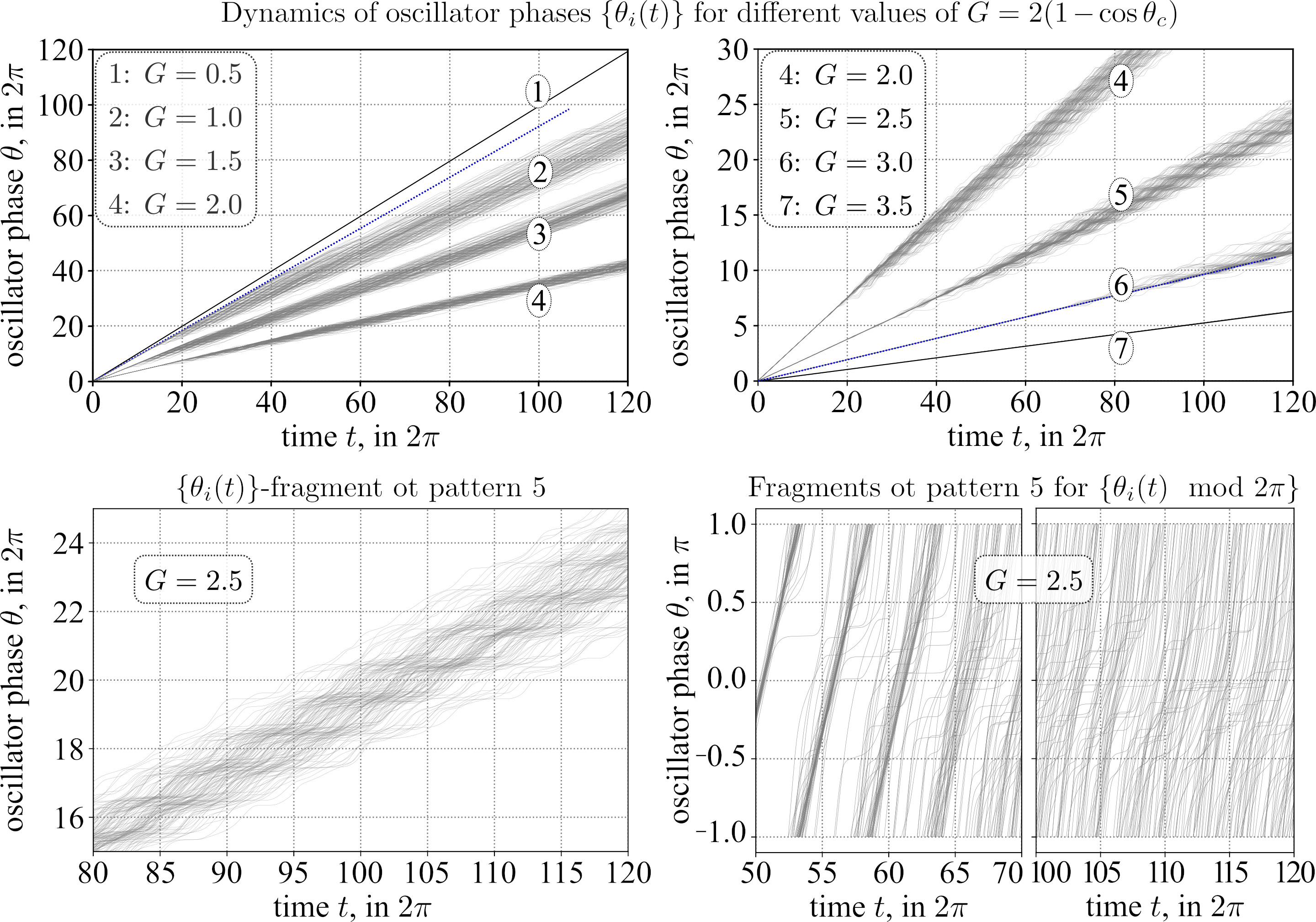}
	\end{center}
	\caption{Dynamics of oscillators for different values of the interaction sector width $G/2$ (Eq.~\ref{eq:G}) in the absence of the synchronization mechanism ($\kappa=0$). The upper row shows the full patterns for 150 selected oscillators, while the lower row presents different fragments of pattern 5, treated either as continuous curves $\{\theta_i(t)\}$  (left plot) or as lines $\{\theta_i(t)\pmod{2\pi}\}$ (with a $-\pi$-shift) confined to the region $[-\pi,\pi]$ (right plot). The blue dotted straight lines (for patterns 2 and 6) indicate the oscillator dynamics assuming their phases could vary synchronously. The other model parameters were set to $\Delta_\theta = 0.2$, $\Delta_\omega = 0$, $\Delta_L = 1.0$, $\Delta_R = 0.5$, and $p=4$.}
	\label{F:3}
\end{figure}

Figure~\ref{F:3} illustrates the dynamics of 150 selected oscillators in the case where all rotation frequencies are identical ($\omega_i = 1$), no synchronization effect is present ($\kappa = 0$), and at the initial time ($t=0$), their phases $\{\theta_i\}$ are randomly distributed within a narrow interval $\theta_i\in (0,\Delta_\theta)$ ($\Delta_\theta\ll \pi$). As an example, we set $\Delta_\theta = 0.2$. In this case, the oscillator dynamics was analyzed for several values of $G$, ranging from 0.5 to 3.5 in increments of 0.5 (Fig.~\ref{F:3}, upper row).

As seen, for small or large values of $G$ (patterns 1  and 7), the synchronous dynamics of the oscillators is, at the very least, a long-lived state. For intermediate values of $G$ (patterns 2--6), numerical simulations demonstrated that this state eventually breaks down, and the mutual orientation of the oscillators becomes random. This is confirmed by the lower right plot, which depicts the time dynamics of $\{\theta_i(t)\mod 2\pi\}$ (shifted by $-\pi$). The loss of synchrony causes the motion of almost every oscillator to be temporarily halted at $\theta_i \pmod{2\pi} = 0$ by its left neighbor (due to $\Delta_L = 1$), resulting in a wavy pattern, as illustrated in the lower left plot. In the lower right plot, this effect is reflected in the numerous horizontal segments of the oscillator trajectories. 

Focusing on patterns 2 and 6, we observe that after the transition to a random phase arrangement, the mean rate of oscillator phase increase,  $d\langle \theta_i(t)\rangle/dt$, deviates by a finite amount from that of its initial quasi-synchronous phase arrangement. For small values of $G$, this shift results in lower rates, while for large values of $G$, it leads to higher rates. This suggests the hypothesis that, in the absence of a synchronization mechanism, the multitude of long-lived states with a synchronous phase arrangement and the multitude of stable states with a random phase arrangement are separated by a finite gap.

\begin{figure}[t]
	\begin{center}
		\includegraphics[width=\textwidth]{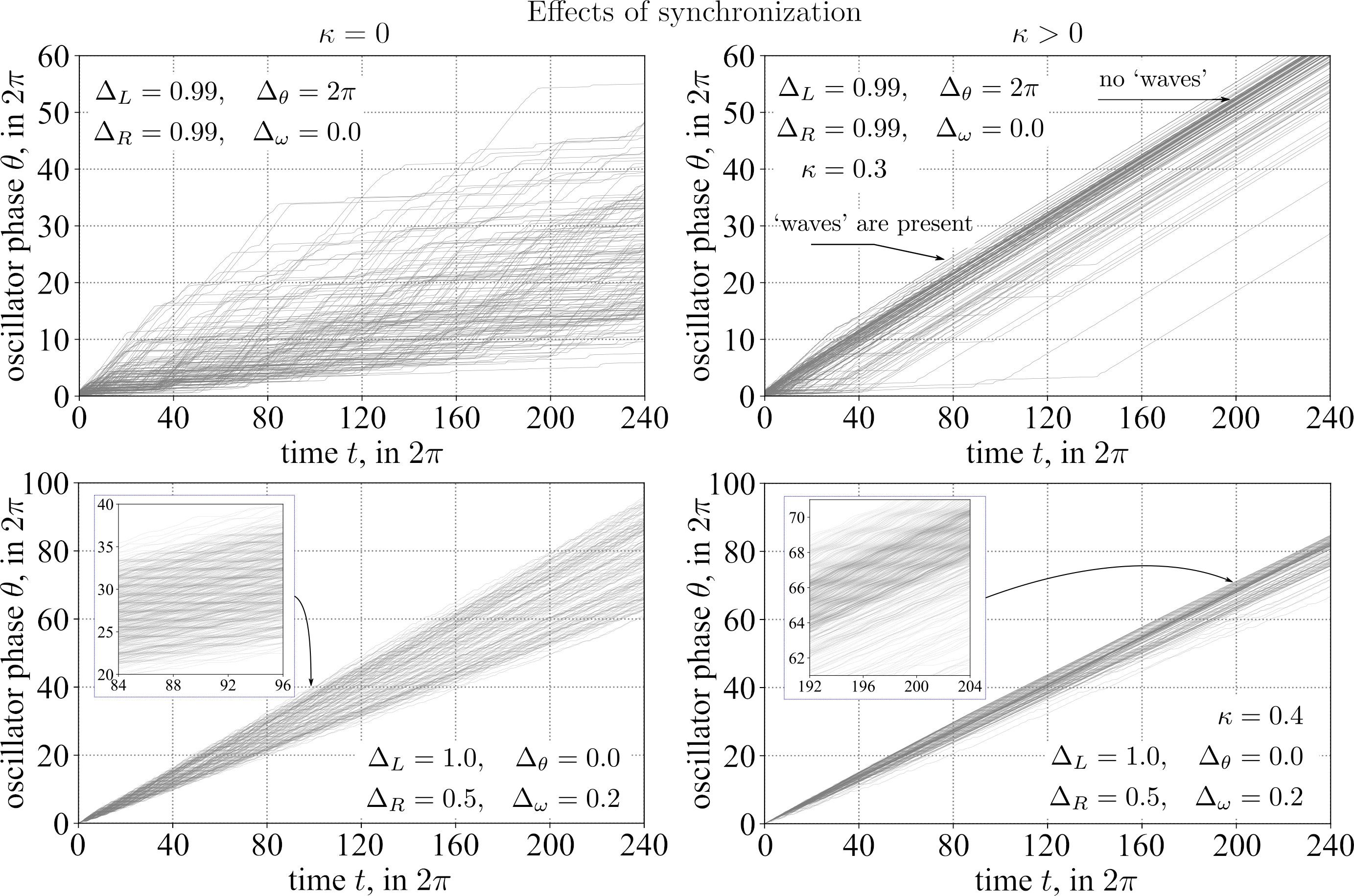}
	\end{center}
	\caption{Illustration of the effect of synchronization on oscillator dynamics near collapse (upper row) and the dynamics of oscillators with different rotation frequencies (lower row). The plots depict trajectory patterns of 150 selected oscillators. The main parameters are shown in the plots, while the other parameters are set to $G = 2$ and $p = 4$.}
	\label{F:4}
\end{figure}

Figure~\ref{F:4} illustrates the effect of synchronization on oscillator dynamics in two limiting cases. First, when the parameters $\Delta_L = \Delta_R = 0.99$ are very close to the critical value of 1, the dynamics of many oscillators can be completely blocked. In this case, the oscillators are assumed to be identical in terms of their rotation frequencies, $\{\theta_i=1\}$,  and the initial distribution of their phases $\{\theta_i\}_{t=0}$ was treated as a set of random variables uniformly distributed in the range $[0,2\pi]$. Second, the oscillators differ in their rotation frequencies, which are random variables $\{\omega_i\}$ uniformly distributed in the interval $\omega_i\in [1-\Delta_\omega, 1+\Delta_\omega]$ with $\Delta_\omega < 1$. In the simulation, we set $\Delta_\omega = 0.2$ while the parameters $\Delta_L = 1$ and $\Delta_R = 0.5$ remained the same as in the previous analysis. In both cases, the width $G/2$ of the interaction sector was set to 1.       

The obtained results for the first case are shown in the upper row. As seen, in the absence of synchronization ($\kappa=0$), the pattern of 150 oscillator trajectories becomes extremely wide, with its size increasing over time (upper left plot). Introducing oscillator synchronization via the proposed mechanism with $\kappa = 0.3$ drastically alters this pattern. After a certain time delay of approximately $40\times2\pi$ units, almost all oscillators synchronize their phases, and their dynamics become nearly uniform, with only minor variations caused by short-term conflicts. After a time interval of about $160\times2\pi$, even these variations vanish.     

The lower row illustrates oscillator synchronization when their rotation frequencies are different. Without synchronization ($\kappa= 0$), the trajectory pattern resembles a bundle of diverging trajectories. The synchronization mechanism with $\kappa=0.4$ significantly contracts these diverging trajectories; numerically, it was found that for $\kappa=0.6$, this pattern appears as a nearly straight line. Despite the substantial contraction, the oscillator phase arrangement remains random, which is reflected in the wavy form of the resulting patterns, as shown in the insets of the lower row plots.

\section*{Conclusion}
We have proposed a novel model of oscillatory chains, $\{e^{i\theta_i}\}$, that generalizes the contour model of Buslaev nets with discrete states. The core idea of the model is a continuous representation of conflicts in system dynamics, modeled as interactions between neighboring oscillators when their phases ${\theta_i}$ approach the points $\theta_i = 0,\pi \pmod{2\pi}$ within interaction sectors of size $\theta_c$. The parameter $\theta_c$ can be interpreted as a measure of the size of vehicle clusters moving along contours. The oscillators are assumed to be capable of synchronizing their dynamics---a process described using terms adapted from the Kuramoto model---which effectively captures for the influence of state discreteness in Buslaev nets.

The derived equation governing the oscillator dynamics $\{\theta_i(t)\}$ incorporates four factors, which is expressed as the product of their corresponding cofactors. First, oscillator $i$ undergoes deceleration due to individual conflicts with its left ($i-1$) and right ($i+1$) neighbors when passing through the points $\theta_i = \pi \pmod{2\pi}$ and $\theta_i = 0 \pmod{2\pi}$, respectively. Conflict resolution may be asymmetric, reflecting differences in the priorities of the left and right oscillators within a given pair $\{i,i+1\}$. Second, oscillator $i$ may accelerate or decelerate as it synchronizes with its left and right neighbors. When the interaction sector $\theta_c$ reaches a value near $\pi/2$, the influence of these four factors on the oscillator dynamics begins to overlap.

Numerical analysis of the dynamics has shown that the system exhibits a range of properties, including both those characteristic of classical Buslaev nets and novel features. The former includes the emergence of metastable synchronization among oscillators, while the latter is exemplified by various phase transitions that occur as the interaction sector varies in size.

This model offers a promising framework for investigating network traffic and its complex dynamics using the proposed approach.

\reftitle{References}

\end{document}